\documentclass[prb,twocolumn,showpacs]{revtex4}

\usepackage{graphicx}
\usepackage{dcolumn}

\newcommand{\beq}{\begin{equation}}
\newcommand{\eeq}{\end{equation}}
\newcommand{\beqa}{\begin{eqnarray}}
\newcommand{\eeqa}{\end{eqnarray}}

\begin{document}

\title{Novel Phase Between Band and Mott Insulators in Two Dimensions}
\author{S. S. Kancharla$^1$ and E. Dagotto$^{1,2}$}
\affiliation{$^1$Materials Science and Technology Division, Oak Ridge
National Laboratory, Oak Ridge, TN 32831 \\ $^2$Department of Physics
and Astronomy, University of Tennessee, Knoxville, TN 37996}
\begin{abstract}
We investigate the ground state phase diagram of the half-filled
repulsive Hubbard model in two dimensions in the presence of a
staggered potential $\Delta$, the so-called ionic Hubbard model, using
cluster dynamical mean field theory. We find that for large Coulomb
repulsion, $U\gg \Delta$, the system is a Mott insulator (MI). For
weak to intermediate values of $\Delta$, on decreasing $U$, the Mott
gap closes at a critical value $U_{c1}(\Delta)$ beyond which a
correlated insulating phase with possible bond order (BO) is
found. Further, this phase undergoes a first-order transition to a
band insulator (BI) at $U_{c2}(\Delta)$ with a finite charge gap at
the transition. For large $\Delta$, there is a direct first-order
transition from a MI to a BI with a single metallic point at the phase
boundary.
\end{abstract}
\pacs{71.10.-w, 71.10.Fd, 71.10.Hf, 71.27.+a, 71.30.+h, 71.45.Lr} 
\date{\today}
\maketitle

The Mott metal-insulator transition is a paradigmatic problem in
condensed matter physics that has been realized in a broad set of
experiments, ranging from the high temperature superconductors to
recently observed phenomena involving cold atoms trapped in optical
lattices. Mott insulators (MI) are characterized by a density of one
electron per unit-cell and strong Coulomb interactions that often
leads to symmetry breaking via antiferromagnetic (AF) order. In
contrast, charge localization in band insulators (BI) occurs when the
number of electrons per unit cell is even, so that the bands are
either full or empty. These two types of insulators can be realized
within a single system, the Ionic Hubbard model (IHM) at half-filling,
by tuning the strength of Coulomb interaction. The BI-MI quantum phase
transition in the IHM has been well studied in the limit of either
one\cite{fabrizio,noack,kampf,aligia} or infinite
dimensions\cite{garg}, but significant uncertainties remain in the
phase diagram. Originally suggested to describe charge-transfer
organic salts\cite{hubbard}, there has been renewed interest in the
model due to potential applications to ferroelectric
perovskites\cite{egami}. Confinement of Fermionic atoms in an optical
lattice to realize the ionic Hubbard model is also conceivable in the
future, because recent experimental progress has allowed for Fermionic
atoms to be cooled well below the degeneracy
temperature\cite{hadzibabic}.

Our objective in this letter is to obtain the ground state phase
diagram of the IHM at half-filling in two dimensions using cluster
dynamical mean field theory (CDMFT)\cite{cdmft}. The IHM is realized
on a bipartite lattice by the assignment of an alternating chemical
potential ($\pm \Delta$) for each sublattice on top of the regular
Hubbard model. The Hamiltonian can be written as \beqa H &=&-t\sum_{i
\in A j\in B} \left[ c_{i\sigma}^{\dagger}c_{j\sigma}+ h.c \right]
+U\sum_{i}n_{i\downarrow}n_{i\uparrow} \nonumber \\ &&+\Delta\sum_{i
\in A}n_{i} - \Delta\sum_{i \in B}n_{i} -\mu\sum_i n_i. \eeqa Here,
$t$ and $U$ denote the hopping amplitude between nearest-neighbor
sites and the onsite Coulomb repulsion, respectively. $+\Delta
(-\Delta)$ denotes the local potential energy for the A(B)
sublattice. We are interested in the case where the combined density
of $A$ and $B$ sublattices is 1 and so we set the chemical potential,
$\mu$, to $U/2$. In the atomic limit ($t=0$) and for $U/2<\Delta$, the
ground state has two electrons on the $B$ sublattice and none on the
$A$ sublattice resulting in charge density wave (CDW) order with a
band gap of $\Delta-U/2$. In the opposite limit where $U/2>\Delta$,
each site is occupied by one electron and a MI is formed with a gap
$U$. Therefore, in the atomic limit at exactly $U_c=2\Delta$ the
system is gapless\cite{fabrizio,noack,kampf,aligia}. This transition
from a BI to a MI is likely to persist when $t\neq0$, but non-trivial
charge fluctuations in the intermediate to strong coupling regime
($U\sim2\Delta$) render static mean-field and perturbative treatments
of the transition invalid.
 
Bosonization calculations for the IHM in one dimension\cite{fabrizio},
valid in the weak coupling regime ($\Delta \ll U \ll t$), find a
transition from a BI to a bond-ordered (BO) insulator at $U=U_c$, where
the charge gap vanishes. The BO phase is characterized by finite
charge and spin gaps. Increasing $U$ further, the spin gap vanishes
at $U=U_s$ in a Kosterlitz-Thouless-type transition leading to a
MI. Numerical studies for long chains using the density matrix
renormalization group (DMRG)\cite{kampf,noack} in the strong coupling
regime find that the charge gap remains finite at the transition point
$U_c$ to the BO phase, although the optical gap does vanish. DMRG is
unable to identify a second transition point towards a MI because
extrapolation to large system sizes in the critical region is a
challenge to numerical methods.

Our key findings for the IHM in two dimensions are the following. A
nonzero staggered chemical potential induces charge density wave (CDW)
order over the entire range of $U$, with the CDW order parameter
approaching zero asymptotically for large $U$. In the large $U$
regime, the system behaves as a MI with AF order. For weak to
intermediate $\Delta$, decreasing the value of $U$, the system
undergoes a transition to a correlated insulating phase suggesting BO
with a closing of the charge gap at the transition point
$U_{c1}(\Delta)$. The transition also manifests itself via an abrupt
increase in the slope of the staggered density, double occupancy, and
staggered magnetization as a function of $U$. Further, the BO phase
undergoes a first-order transition to a BI at a lower critical
coupling $U_{c2}(\Delta)$ where the charge gap remains finite. AF
order is found only in the MI and BO phases. For $\Delta\sim 4.5t$,
the system undergoes a direct first-order transition from the BI to
the MI, with a finite charge gap everywhere.

CDMFT is a non-perturbative technique where the full many-body problem
is mapped onto local degrees of freedom treated exactly within a
finite cluster that is embedded in a self-consistent bath. It is a
natural generalization of single-site DMFT to incorporate spatial
correlations. The method has passed rigorous tests for the 1D Hubbard
model where it compares well to exact solutions\cite{cdmftapp}. CDMFT
has also been used to elucidate various aspects of the phase diagram
of the cuprates, including the pseudogap\cite{kyung} and
superconducting phases within the 2D Hubbard model\cite{sarma}.

Using CDMFT, the IHM on the infinite lattice in two dimensions reduces
to the cluster-bath Hamiltonian below, that is subject to a self-consistency
condition:  \beqa H&=&\sum_{\langle\mu\nu\rangle\sigma}
t_{\mu\nu}c_{\mu\sigma}^{\dagger}c_{\nu\sigma} + U\sum_\mu
n_{\mu\uparrow}n_{\mu\downarrow} +\Delta\sum_{\mu} (-1)^\mu n_{\mu}
\nonumber \\
&+&\sum_{m\sigma}\epsilon_{m\sigma}a_{m\sigma}^{\dagger}a_{m\sigma}+\sum_{m\mu\sigma}V_{m\mu\sigma}(a_{m\sigma}^{\dagger}c_{\mu\sigma}+h.c).
\label{clusterham}
\eeqa $\mu,\nu=1,\cdot\cdot\cdot, N_c$ denote indices labelling
the cluster sites and $m=1,\cdot\cdot \cdot,N_b$ represent those in
the bath. The self-consistent calculation proceeds by an initial guess
for the cluster-bath hybridization $V_{m\mu\sigma}$ and the bath site
energies $\epsilon_{m\sigma}$ to obtain the cluster Green's function
$G_c^{\mu\nu}$. Applying the Dyson's equation,
$\Sigma_c=G_{0c}^{-1}-G_{c}^{-1}$, where $G_{0c}$ denotes the
non-interacting Green's function for the Hamiltonian in
Eq.~\ref{clusterham}, the cluster self-energy is obtained.  $\Sigma_c$ is
then used in the self-consistency condition below to determine a new
$G_{0c}$:  \beq G_{0c}^{-1}(z)-\Sigma_c(z)=\left[\frac{N_c}{4\pi^2}
\int dK \frac{1}{z+\mu-t(K)-\Sigma_c(z)}\right]^{-1}.
\label{selfcon}
\eeq Here, $K$ denotes a momentum vector in the reduced Brillouin zone
of the cluster superlattice and $z=i\omega_n$ is the Fermionic
Mastubara frequency. From $G_{0c}$ in Eq.~\ref{selfcon} a new set of
$V_{m\mu\sigma}$ and $\epsilon_{m\sigma}$'s is generated closing an
iterative loop. In this work, the Lanczos method is used to solve the
cluster-bath Hamiltonian. The cluster size is fixed to $N_c=4$ and the
bath size to $N_b=8$. The Lanczos method can access both the strong
and weak coupling regimes with equal ease and it is well suited to
compute dynamical quantities directly in real frequency. Rotational
symmetries of the cluster on a square lattice, together with
particle-hole symmetry at half-filling, reduce the number of bath
parameters significantly. These bath parameters can depend on spin,
allowing for symmetry breaking solutions.  For details of the method,
we refer to earlier work\cite{cdmft,cdmftapp}.
\begin{figure}[htb]
\begin{center}
\includegraphics[width=6.0cm,angle=-0] {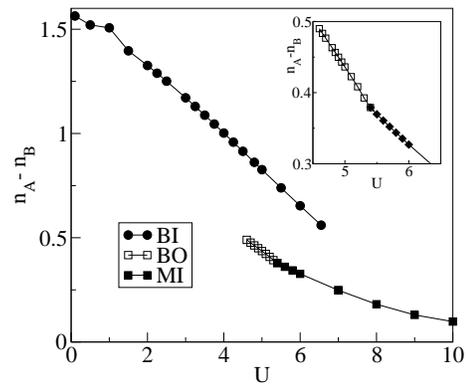}
\end{center}
\caption{Staggered charge density as a function of $U$ in the BI, MI,
and BO regions for $\Delta=2t$. Inset shows an enlarged view of the
abrupt change of slope at the MI-BO boundary.}
\label{staggered-density}
\end{figure}
\begin{figure}[htb]
\begin{center}
\includegraphics[width=8.0cm,angle=-0] {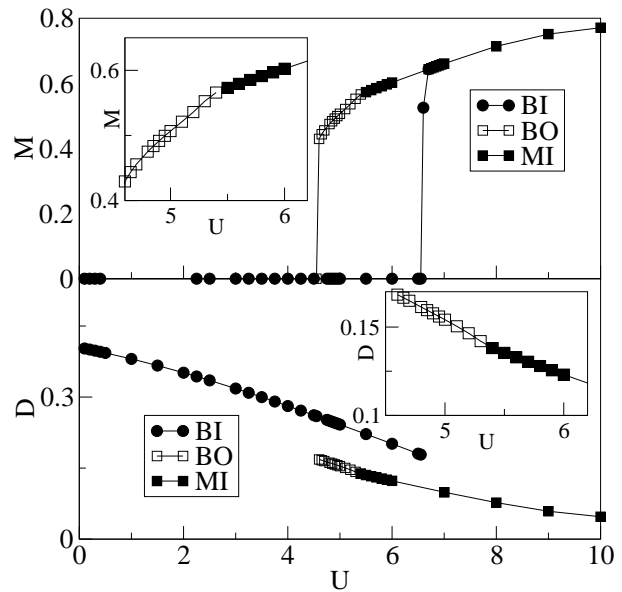}
\end{center}
\caption{(a) Staggered magnetization, 
and (b) double occupancy 
as a function of $U$, for $\Delta=2t$.}
\label{magdoub}
\end{figure}

We analyze our results by starting with Fig.~\ref{staggered-density},
which shows the density difference between sublattices A and B as a
function of $U$, for $\Delta$=$2t$. Clearly, the system presents 
long-range charge density wave (CDW) order over the entire range of
$U$. The CDW order parameter asymptotically approaches zero for
increasing $U$ but remains finite everywhere. Starting from the MI
phase at large $U$, the staggered charge density increases smoothly as
$U$ is reduced, but at $U$=$U_{c1}$=$5.4t$ there is an abrupt
increase in the slope. This anomaly is clearly identified in the inset
and it is believed to be a signature of the transition to a Bond-Ordered (BO)
phase. This result is similar to observations in earlier DMRG
calculations for the one dimensional 
IHM\cite{kampf,noack}. The BO
phase extends up to $U$=$U_{c2}$=$4.6t$, where there is a first-order
transition to a BI signalled by a jump in the staggered
density. Starting from $U=0$, the BI shows a continous decrease in the
staggered density and persists up to $U$=$6.55t$. Consequently, we have
the coexistence of a BI and BO for $4.6t<U<5.4t$ and a BI and MI for
$5.4t<U<6.55t$. We computed the total energy in the coexistence region
(not shown) and find that the BI solution for the CDMFT equations is
always higher in energy than the BO or the MI phase for the same $U$.
\begin{figure}[ht]
\begin{center}
\includegraphics[width=6.0cm,angle=-0] {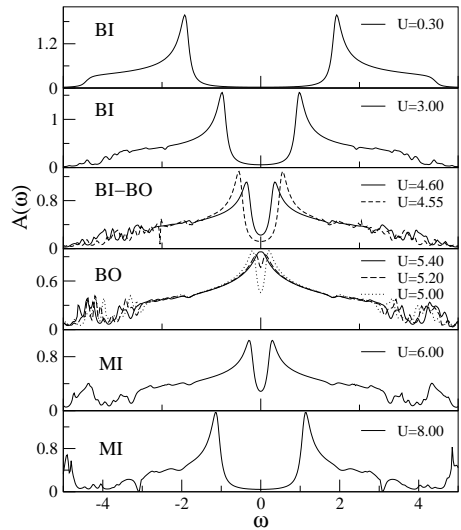}
\end{center}
\caption{Local density of states as a function of $U$ for $\Delta=2t$
shown with a broadening factor of $0.02t$.}
\label{spectra}
\end{figure}

To further characterize the nature of the transition let us consider
the staggered magnetization, $M$=$n_\uparrow-n_\downarrow$, and the
double occupancy, $D$=$\langle n_{\uparrow}n_\downarrow\rangle$, as a
function of $U$ for $\Delta$=$2t$ shown in Fig.~\ref{magdoub}. With
reducing $U$ both these quantities show an abrupt increase in slope at
the same transition point $U$=$U_{c1}$=$5.4t$ (see inset) where the BO
insulator emerges from the MI. Both $M$ and $D$ present a jump at the
first-order transition from the BO phase to a BI. The staggered
magnetization abruptly goes to zero upon entering the the BI phase
signifying an absence of AF order.  The double occupancy increases
monotonically with reducing $U$ in the BI state due to the increasing
tendency towards CDW order.

In Fig.~\ref{spectra}, results for the local density of states (LDOS)
as a function of $U$ for $\Delta=2t$ are presented.  Note that the
LDOS shows particle-hole symmetry consistent with half-filling because
it is averaged over the A and B sublattices. For $U$ equal to the
bandwidth of $8t$, the spectra show lower and upper Hubbard bands
consistent with a MI. In addition, sharp coherent peaks are observed
(absent in single-site DMFT). They correspond to the motion of a
carrier in an AF background and define the Mott gap.  As $U$ is
lowered, the Mott gap reduces its value until it closes at the
transition point $U=5.4t$, which manifested itself earlier as an
abrupt change in slope for the staggered density, double occupancy,
and staggered magnetization. With further reduction in $U$, the trend
for the Mott gap is reversed, with an increase in the gap in the BO
phase. Further, at $U=4.55$ there is an abrupt jump in the gap as
compared to $U=4.60$ signalling a transition to a BI, which also shows
an increasing gap with decreasing $U$. For $\Delta>4.5t$, the BI-MI
transition has no intervening phase and it is accompanied by the
closing of the charge gap. This transition scenario is reminiscent of
the phase diagram of the extended Hubbard model in one dimension,
where an intermediate phase with bond order is stabilized between the
CDW and SDW phases only for intermediate strength of Coulomb repulsion
$U$ when the nearest neighbor Coulomb repulsion $V$ is
varied\cite{nakamura}.
\begin{figure}[h]
\begin{center}
\includegraphics[width=5.5cm,angle=-0] {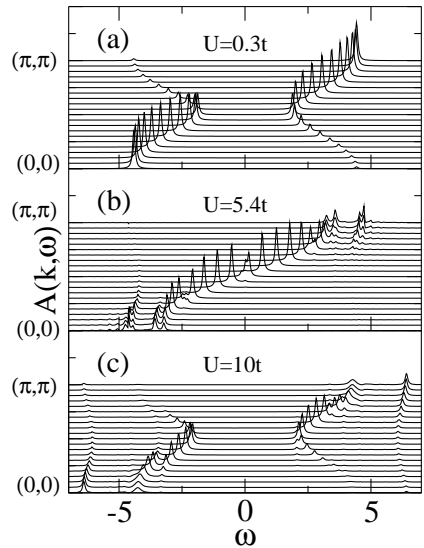}
\end{center}
\caption{Momentum dependent density-of-states along the $(0,0)$ to
$(\pi,\pi)$ direction for (a) the BI, (b) the phase
boundary of the MI-BO transition, and (c) the MI for $\Delta=2t$.}
\label{00pipi}
\end{figure}

To characterize the distinct phases observed, we have computed the
momentum-dependent density-of-states using the following relation
between the the lattice Green's function and the cluster self-energy:
\beq G({\bf k},z)=\frac{1}{N_c}\sum_{\mu\nu}e^{i{\bf
k}\cdot(\mu-\nu)}\left[\frac{1}{z+\mu-t({\bf
k})-\Sigma(z)}\right]_{\mu\nu}. \eeq In Fig.~\ref{00pipi}, we show the
evolution of the density of states, $A({\bf k},\omega)$, in the first
quadrant of the Brillouin zone, from $(0,0)$ to $(\pi,\pi)$. At large
$U$ (Fig.~\ref{00pipi}(c)) sharp coherent peaks that define the Mott
gap are observed. In addition, the lower and upper Hubbard bands are
present, as well as high energy features separated from them by
$\Delta$.  The minimum of the gap is found at ${\bf
k}$=$(\pi/2,\pi/2)$ where the low energy features disperse towards
zero frequency, while the maximum of the gap is found at ${\bf
k}$=$(\pi,\pi)$ or $(0,0)$. For small $U$ (Fig.~\ref{00pipi}(a)),
where we have a BI, a dispersion characteristic of AF order is
observed but with the notable absence of Hubbard bands. By contrast,
for intermediate $U$, on the phase boundary of the MI-BO transition at
$U=5.4t$, we found a metallic density of states with distinct higher
energy bands at $(\pi,0)$ and $(0,\pi)$ that emerge from spontaneous
dimerization and persist into the BO phase. Therefore, the BO phase,
although similar to the BI in the behavior of the charge gap as a
function of $U$, shows a distinct signature of strong correlations.
The metallic peak itself splits to open a gap right below the phase
boundary.
\begin{figure}[htb]
\begin{center}
\includegraphics[width=5.0cm,angle=-0] {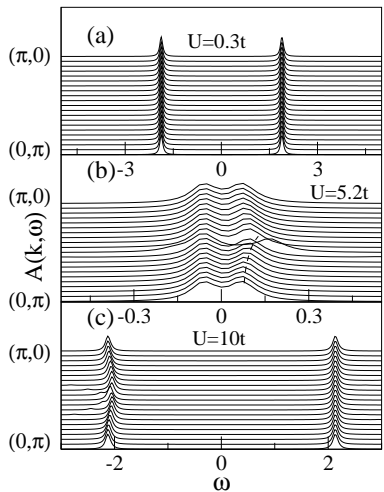}
\end{center}
\caption{Momentum dependent density-of-states along the $(0,\pi)$ to $(\pi,0)$ direction for $\Delta=2t$.}
\label{0pipi0}
\end{figure}
\begin{figure}[ht]
\begin{center}
\includegraphics[width=5.8cm,angle=-0] {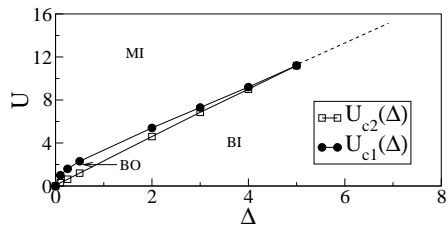}
\end{center}
\caption{Proposed CDMFT phase diagram for the 2D IHM.}
\label{phasediag}
\end{figure}

In Fig.~\ref{0pipi0}, $A({\bf k},\omega)$ is shown, from $(0,\pi)$ to
$(\pi,0)$. The uncorrelated BI at $U=0.3t$ (Fig.~\ref{0pipi0}(a))
shows no dispersion of the low lying features and the gap becomes
independent of momentum. For the MI phase at $U=10t$
(Fig.~\ref{0pipi0}(c)) we see an inward dispersion in the occupied
band at ${\bf k}$=$(\pi/2,\pi/2)$ where the gap is a minimum,
consistent with commensurate AF fluctuations. In contrast to this
result, Fig.~\ref{0pipi0}(b) for $U=5.2t$ shows a maximum of the
charge gap at ${\bf k}$=$(\pi/2,\pi/2)$ while the minimum lies at
${\bf k}$=$(0,\pi)$ or $(\pi,0)$, consistent with the spontaneously
dimerized BO phase with incommensurate order.

To summarize, we have studied the quantum phase transition between a
BI and a MI in the two dimensional IHM using CDMFT. CDW order persists
throughout the entire range of $U$. For $\Delta<\Delta_c\sim4.5t$, the
large $U$ Mott insulator state undergoes a transition to a correlated
insulating phase BO with reducing $U$, accompanied by the closing of
the charge gap at the transition point $U_{c1}(\Delta)$. The BO phase
undergoes a first-order transition to a BI at $U_{c2}(\Delta)$ upon
further reduction of $U$. As shown in the phase diagram depicted in
Fig.~\ref{phasediag}, the width of the BO phase shrinks as both small
and large $\Delta$ are approached. For $\Delta>\Delta_c$, a direct
first-order MI-BI transition with no intervening BO phase is observed.
AF order is found only in the MI and BO phases. Our results differ
significantly from earlier ones that used the single-site DMFT
technique\cite{garg} where a paramagnetic solution must be enforced
even though the ground state is AF. Single-site DMFT misses spatial
correlations crucial to the physics of this model. Moreover, the
single-site DMFT finds a finite metallic region between the BI and MI
whereas we find a single metallic point at the MI-BO
($\Delta<\Delta_c$) or MI-BI ($\Delta>\Delta_c$) transitions. Longer
range interactions could give rise to a metallic phase between the
insulators\cite{poilblanc}. Our results are fairly similar to those
obtained for this model using DMRG \cite{noack,kampf} in one
dimension. An important difference is that at the metallic point we
observe a closing of the charge gap whereas DMRG finds only a closing
of the optical gap, with the charge gap remaining finite over the
entire range of $U$. The two-transition scenario moving from the BI to
the MI was observed earlier in the weak coupling regime using
Bosonization\cite{fabrizio} for this model in one dimension. We have
shown that a similar transition persists in two dimensions, even in
the strong coupling regime when $\Delta$ is comparable to half the
bandwidth. Modulation of the local potential with an alternating value
could be realized in an two dimensional optical lattice loaded with
ultracold Fermionic atoms. Noise correlations in time-of-flight images
can be used to distinguish between the AF, BI and BO
phases\cite{greiner}. Energy gaps can be obtained, in principle, using
rf spectroscopy\cite{chin}. We expect our results, predicting a novel
correlated insulating phase with bond order, to motivate such
experiments to verify our finding.

S.S.K. is supported by the LDRD program at Oak Ridge National
Laboratory. E.D. acknowledges support by the NSF Grant
No. DMR-0454504.

\end{document}